\documentclass[a4paper]{jpconf}
\usepackage{graphicx}
\usepackage{helvet}
\begin{document}
\title{Quantum Scattering in an Optical Collider for Ultracold Atoms}

\author{Ryan Thomas, Matthew Chilcott, Craig Chisholm, Amita B. Deb, Milena Horvath, Bianca J. Sawyer, and Niels Kj{\ae}rgaard}

\address{Department of Physics, QSO---Centre for Quantum Science, and Dodd-Walls Centre for Photonic and Quantum Technologies, University of Otago, Dunedin, New Zealand}

\ead{niels.kjaergaard@otago.ac.nz}

\begin{abstract}
We report on experiments investigating the collisional properties of atoms at ultralow collision energies using an all-optical atom collider.  By using a pair of optical tweezers, we can manipulate two ultracold atom clouds and collide them together at energies up to three orders of magnitude larger than their thermal energy.  Our experiments measure the scattering of $\rm ^{87}Rb$, $\rm ^{40}K$, and $\rm ^{40}K$-$\rm ^{87}Rb$ collisions.  The versatility of our collider allows us to probe both shape resonances and Feshbach resonances in any partial wave. As examples, we present experiments demonstrating p-wave scattering with indistinguishable fermions, inelastic scattering at non-zero energies near a homonuclear Feshbach resonance, and partial wave interference in heteronuclear collisions.
\end{abstract}

\section{Introduction}
The scattering experiment provides one of the most powerful ways of studying physical systems, epitomised by Geiger and Marsden's angular-resolved measurements of $\alpha$-particles impinging on a gold foil \cite{Geiger1909} which led to Rutherford's atomic model \cite{Rutherford1911}. Throughout the century that has passed since this seminal work, scattering experiments have remained a method of choice to probe atoms and molecules \cite{Kleinpoppen2013}.

\begin{figure}
\begin{center}
\includegraphics[width=\textwidth]{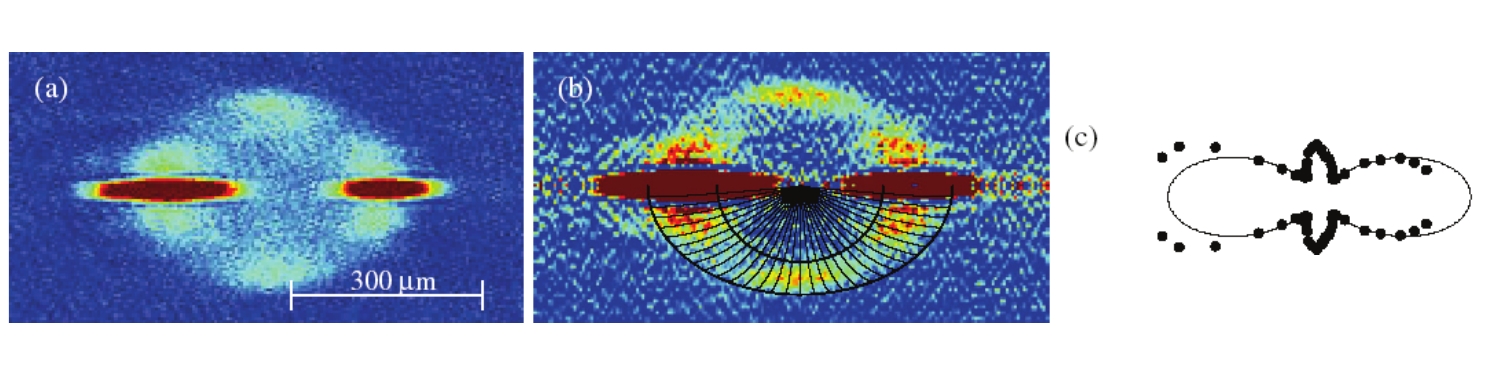}
\caption{Observation of s and d-wave inteference in the collision of two atomic clouds of $\rm ^{87}Rb$.  (a) Unprocessed absorption image.  (b) Absorption image after Abel-inversion with curved angular bins drawn (accounting for expansion in an anisotropic harmonic trap.  (c) Angular distribution of scattered particles extracted from (b).  Figure adapted from Ref.~\cite{Kjaergaard2004} {\tiny\textsf{\copyright}~IOP Publishing \& Deutsche Physikalische Gesellschaft. CC BY-NC-SA}}
\label{fg:Fig1}
\end{center}
\end{figure}

With the advent of techniques to trap and cool atomic gases to the nanokelvin regime, possibilities have opened for capturing quantum scattering in its purest form.  While most experiments involving atomic collisions at ultralow energies rely on thermal ensembles without a well-defined collision axis \cite{Weiner1999}, a small number of experiments have emerged which use a truly collider-like geometry.  In an early example, Gibble \textit{et~al.} \cite{Gibble1995} used a cesium fountain to launch cold clouds of atoms in an upward direction, where they would collide --- a technique that was later refined to measure the p-wave threshold of cesium \cite{Legere1998}. The atomic fountain experiments also subsequently introduced an interferometric method that considered the effect of collisions on atoms in a superposition of internal clock states \cite{Hart2007} and this was used to locate Feshbach resonances and measure their shift with magnetic field \cite{Gensemer2012}. However, as this technique measures the relative scattering phase shift between two clock states in cesium, it cannot be readily extended to arbitrary species of atoms in arbitrary internal quantum states. An alternative scheme was provided by magnetic colliders \cite{Thomas2004a}. Here a magnetic double-well trap containing two separate ultracold samples of atoms would be continuously converted into a single-well potential. The atoms would then collide along a single direction at specific energies defined by the ramp parameters for the double-to-single well conversion. This technique was used to observe collisions of $\rm ^{87}Rb$ atoms at energies up to 1.2 mK \cite{Thomas2004,Buggle2004,Mellish2007}, measured in units of the Boltzmann constant.  Absorption imaging was used to directly observe the scattering halo as shown in Fig.~\ref{fg:Fig1}.
By using standard image processing techniques such as Abel-inversion, experiments demonstrated s and d-wave interference for the case of indistinguishable bosons \cite{Thomas2004,Buggle2004} and s, p, and d-wave interference for the case of distinguishable bosons\footnote{Perhaps more accurately these should be referred to as identical bosons (atoms of the same isotope) in distinguishable internal states (atoms prepared in different hyperfine states). Without diverting into a discussion on the concepts of identity and indistinguishability \cite{DEMUYNCK1975} we shall just note that identical particles in distinguishable internal states give rise to the same expression as for nonidentical particles in regards to their elastic scattering cross section \cite{Walraven2014} (see also \cite{Feynman1989}). For the purpose of our differential scattering measurements, identical particles prepared in different internal states can effectively be considered as distinguishable or unlike atoms. Notwithstanding that, it should also be noted that the two situations --- collisions between identical particles in distinguishable states versus collisions between nonidentical particles --- are in fact subtly different: the former can generate spin-spin entanglement, the latter cannot \cite{Lamata2006}.} \cite{Mellish2007}.  While the magnetic collider could be used with any magnetically trappable states, this roughly halved the number of possible states that could be investigated.  Furthermore, because the magnetic field could not be turned off sufficiently fast, collisions occurred while in trap, which complicated the analysis of the images since scattered particles would acquire curved trajectories in the anisotropic harmonic potential (see Fig. 1b).  Finally, magnetically tunable Feshbach resonances, which are of prime interest to the ultracold atomic physics community \cite{FeshbachReview}, could not be investigated because the bias field for the magnetic collider was fixed. An important observation that was made in these early experiments using magnetic colliders was that unequal sizes (in terms of particle number) of the incoming clouds gave rise to asymmetric halos (see Fig.~1c where left-right asymmetry is apparent).
If scattering only resulted from binary events the back-to-back nature of elastic scattering would lead to symmetric halos: every particle going out in a particular direction would have a partner going out in the exact opposite direction. Asymmetric halos then strongly indicate that multiple scattering needs to be taken into account to properly analyse the halos when obtaining differential cross sections.

In response to the limitations of magnetic colliders, we developed a laser-based collider \cite{Rakonjac2012a} by means of steerable optical tweezers \cite{Roberts2014}.  The use of optical fields means that atoms in any internal state can be explored in the collider, that collisions can take place in free space since the lasers can be switched off on a nanosecond timescale, and, crucially, magnetic bias fields can be applied to investigate magnetic Feshbach resonances. As part of our collider research program we have also developed analysis methods where effects of multiple scattering are taken into account via direct simulation Monte Carlo (DSMC) modelling \cite{Wade2011}. In these proceedings we present a summary of collision experiments that we have undertaken using our second-generation optical collider.

\section{Collisions of indistinguishable fermions}
It is well-known that collisions of indistinguishable bosons or fermions are dramatically different: while the wavefunction for bosons must have even parity, the wavefunction for fermions must have odd parity. For fermions this anti-symmetrization requirement means that scattering cannot occur into angles at $90^\circ$ to the collision axis -- a so-called forbidden region. Any scattering must be via odd partial waves and these all have a node at  $\theta=90^\circ$, where  $\theta$ is the angle with respect to the collision axis. At low energies and for short-ranged interactions, all scattering is via the $\ell=1$ partial wave due to Wigner threshold suppresion \cite{Wigner1948} and the wavefunction for outgoing fermions in the center-of-mass frame must have a p-wave angular distribution proportional to $\cos^2\theta$.

\begin{figure}[t!]
\includegraphics[width=3.5in]{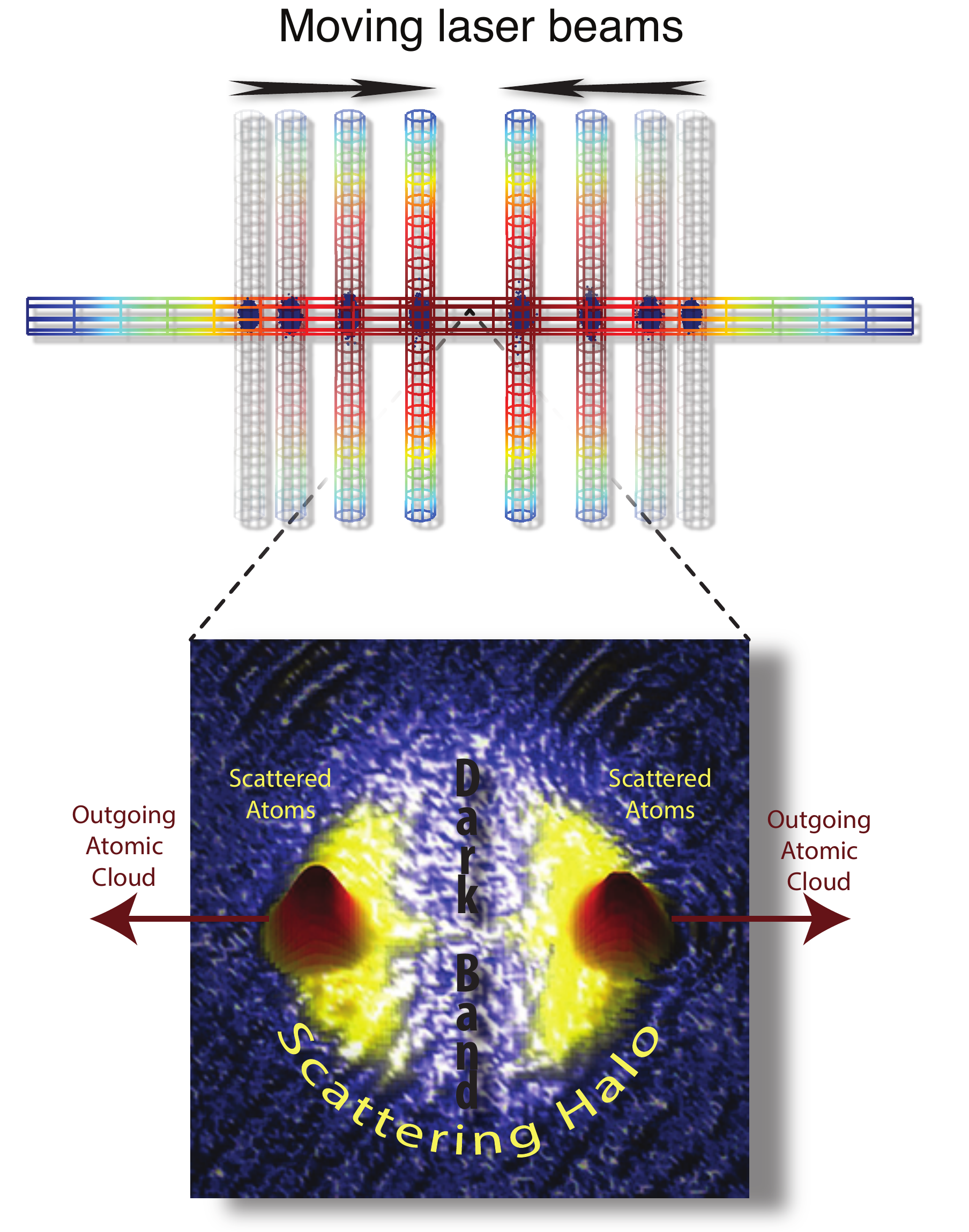}
\hspace{2pc}
\begin{minipage}[b]{14pc}
\caption{Collisions of fermionic $\rm^{40}K$ atoms using an optical collider.  Coloured grids indicate laser beams which trap and steer the potassium atoms.  The lower image is a surface plot of an absorption image for a collision experiment conducted at an energy of 150 $\mu$K, which clearly shows the absence --- a dark band --- of potassium atoms at angles of $90^\circ$ with respect to the collision axis.}
\label{fg:Fig2}
\end{minipage}
\end{figure}

Using our optical collider, we accelerated two clouds of fermionic $\rm ^{40}K$ atoms each in the $|F=\frac{9}{2},m_F=\frac{9}{2}\rangle$ state towards each other at collision energies from 50 to 1800 $\mu$K \cite{Thomas2016}. Figure~\ref{fg:Fig2} shows a post-collision image of the outgoing clouds accompanied by a halo of scattered particles for an experiment conducted at an energy of 150 $\mu$K.
%\begin{figure}
%\begin{center}
%\includegraphics[width=3.0in]{fig1.pdf}
%\caption{Collisions of $\rm^{40}K$ atoms using an optical collider.  Coloured grids indicate laser beams which trap and steer the potassium atoms.  The lower image is an absorption image at an energy of 150 $\mu$K, which clearly shows the absence of potassium atoms at angles of $90^\circ$ with respect to the collision axis.}
%\label{fg:Fig2}
%\end{center}
%\end{figure}
The collision scatters the indistinguishable fermionic potassium atoms into a p-wave halo with a distinct region where scattered atoms are absent at $90^\circ$ to the collision axis.  While one na\"{i}vely would expect that this forbidden region should be present at all energies due to the underlying symmetry of the wavefunction, we actually find that for a certain energy range significant numbers of atoms are scattered at $90^\circ$ angles.  The cause of this apparent violation of anti-symmetrization is a combination of a high density of atoms and a large collision cross section due to a broad p-wave shape resonance at $\approx 350$ $\mu$K.  In this collisionally opaque regime, multiple scattering events become very important.  Near the peak of the resonance, higher-order scattering events are most likely to result in p-wave collision halos that are slightly rotated from the collision axis, and an average over all these rotations gives rise to an apparent isotropic scattering halo that is superimposed on the expected p-wave halo \cite{Thomas2016}.  At energies well above the shape resonance, multiple scattering events give rise to an enhancement in the number of atoms scattered close to the collision axis. Because of the pristine p-wave nature of the underlying binary scattering pattern, this provides an ideal setting to test and verify our DSMC code for modelling multiple scattering dynamics.

\section{Measuring inelastic scattering near a Feshbach resonance}
\begin{figure}[b!]
\begin{center}
\includegraphics[width=0.75\textwidth]{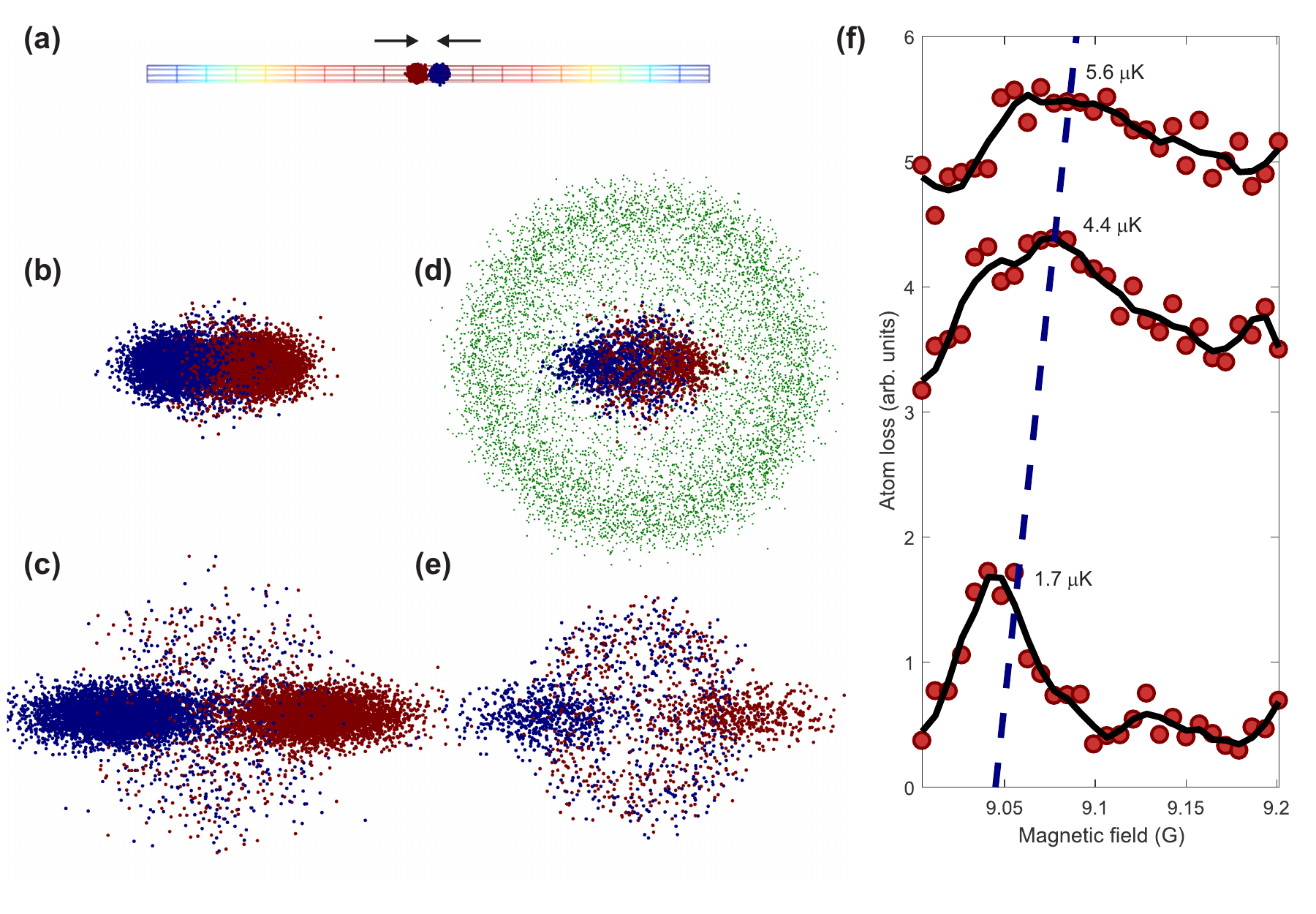}
\caption{\label{fg:Fig3} Collisions of $\rm ^{87}Rb$ atoms in two different states.  {(a)} Two atom clouds (red and blue circles) collide while still partially trapped inside a horizontal laser beam (coloured grid). {(b),(c)}  When the clouds collide away from the resonance, only a few atoms are elastically scattered and none are inelastically scattered.  After a sufficient expansion time, we count the remaining atoms. {(d),(e)} When the clouds collide on resonance, many atoms are inelastically scattered into undetected states (green) with higher kinetic energies and are lost.  After a sufficient amount of time, we count the remaining atoms. {(f)} Measurements of inelastic loss as a function of magnetic field at energies 1.7 $\mu$K, 4.4 $\mu$K, and 5.6 $\mu$K as annotated.  Red circles are data points, black line is a smoothed version of the data, and the dashed blue line is the expected resonance position.}
\end{center}
\end{figure}

As a second demonstration of our collider, we considered a narrow Feshbach resonance of $\rm ^{87}Rb$ centered at 9.045 G (at threshold \cite{Sawyer2017}) between the internal states $|2,0\rangle$ and $|1,1\rangle$ for which we measured inelastic scattering as the resonance was magnetically tuned to lie above threshold \cite{Horvath2017}.  Elastic scattering for this system is weak, producing faint scattering halos, but the inelastic loss into undetected and higher kinetic energy states is significant.  Rather than counting the number of atoms in the collision halo, we count the atoms that remain in the original state as illustrated in Fig.~\ref{fg:Fig3}.
%\begin{figure}[h]
%\includegraphics[width=3.5in]{Figure3.pdf}
%\hspace{2pc}
%\begin{minipage}[b]{14pc}
%\caption{\label{fg:Fig3} Collisions of Rb atoms in two different states.  \textbf{(a)} Two atom clouds (red and blue circles) collide while still partially trapped inside a horizontal laser beam (coloured grid).  \textbf{(b),(c)}  When the clouds collide away from the resonance, only a few atoms are elastically scattered and none are inelastically scattered.  After a sufficient expansion time, we count the remaining atoms.  \textbf{(d),(e)} When the clouds collide on resonance, many atoms are inelastically scattered into undetected states (green) with higher kinetic energies and are lost.  After a sufficient amount of time, we count the remaining atoms.}
%\end{minipage}
%\end{figure}
Using this method, we can measure inelastic loss over a range of collision energies, although energy-broadening mechanisms inherent to our collider limit the signal-to-noise at moderate collision energies.  As can be seen in Fig.~\ref{fg:Fig3}f, the resonance profile becomes noticeably broader as the collision energy increases, and the signal contrast becomes weaker.  However, we have found that the narrow resonance imprints itself on the spatial profile of the atomic clouds, and we can use this imprint to accurately determine the resonance position over a range of collision energies even when the observed loss is within the noise level of our measurements \cite{Horvath2017}.

\section{Heteronuclear collisions and future directions}
\begin{figure}[t!]
\begin{center}
\includegraphics[width=0.8\textwidth]{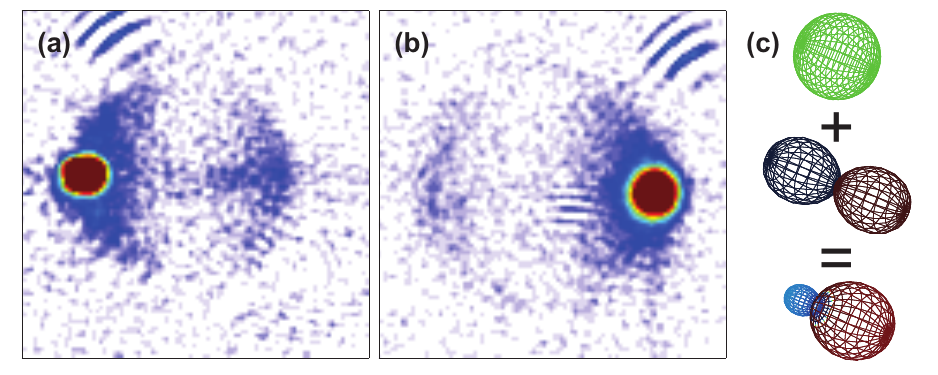}
\caption{Absorption images of {(a)} $\rm ^{40}K$ and {(b)} $\rm ^{87}Rb$ after a collision at 152 $\mu$K near a p-wave Feshbach resonance.  {(c)} Illustration of partial wave interference.  An s-wave amplitude adds to a p-wave amplitude to give an intermediate form.  Radial coordinate indicates magnitude of the scattering amplitude, while the colour indicates the phase.}
\label{fg:Fig4}
\end{center}
\end{figure}

Finally, we demonstrate heteronuclear collisions between $\rm ^{87}Rb$ and $\rm ^{40}K$ atoms in Fig.~\ref{fg:Fig4}.
%\begin{figure}[h]
%\includegraphics[width=3.5in]{KRbPwave.pdf}
%\hspace{2pc}
%\begin{minipage}[b]{14pc}
%\caption{\label{fg:Fig4} Absorption images of \textbf{(a)} $\rm ^{40}K$ and \textbf{(b)} $\rm ^{87}Rb$ after a collision at 152 $\mu$K near a p-wave Feshbach resonance.}
%\end{minipage}
%\end{figure}
Since the atoms comprising the collision pair are distinguishable, all partial waves --- even and odd --- can contribute to the scattering process.  For this particular collision energy (152~$\mu$K), only s and p-wave contributions are relevant.  In Fig.~\ref{fg:Fig4}, the p-wave component has been enhanced by a nearby p-wave Feshbach resonance.  The images clearly show the interference between the s and p-wave processes, exemplified by the anisotropic scattering that is asymmetric about the collision point in the middle of the image.  By analysing the shape of the scattering halo, we can extract both the s and p-wave scattering phase shifts.  Furthermore, by measuring the shape of the halo as a function of magnetic field, we can extract the resonant phase change across the Feshbach resonance.

Future experiments using the optical collider will investigate homonuclear and heteronuclear collisions near s, p, and d-wave Feshbach resonances, where we can accurately measure the threshold behaviour of these resonances and also refine models of the inter-atomic potential.  We will also investigate further shape resonances, such as an $\ell=6$ shape resonance in $\rm ^{87}Rb$ and an $\ell=4$ shape resonance in $\rm ^{40}K$-$\rm ^{87}Rb$ collisions.  We can also associate molecules near a Feshbach resonance and measure atom-dimer scattering properties \cite{Croft2017}.
\ack
We thank Ana Rakonjac, Julia Fekete, Thomas McKellar, and Kris Roberts for their role in constructing the optical collider. We acknowledge fruitful discussions with Eite Tiesinga and Blair Blakie on the matters of resonant scattering and DSMC modeling, respectively.

\section*{References}
\providecommand{\newblock}{}

%\section*{References}
%\begin{thebibliography}{9}
%\bibitem{iopartnum} IOP Publishing is to grateful Mark A Caprio, Center for Theoretical Physics, Yale University, for permission to include the {\tt iopart-num} \BibTeX package (version 2.0, December 21, 2006) with  this documentation. Updates and new releases of {\tt iopart-num} can be found on \verb"www.ctan.org" (CTAN).
%\end{thebibliography}

\end{document}